\documentclass[twocolumn,prl,showpacs,superscriptaddress]{revtex4}
\pdfoutput=1
\usepackage{setspace}
\usepackage{amsfonts,amsmath,amssymb}
\usepackage{txfonts}
\usepackage{graphicx, color}
\usepackage{hyperref}

\begin{document}

\title{Nonequilibrium dynamics of one-dimensional hard-core anyons following a quench:  Complete relaxation of one-body observables}

\author{Tod M. Wright}
\email{todw@physics.uq.edu.au}
\affiliation{The University of Queensland, School of Mathematics and Physics, Brisbane, Queensland 4072, Australia}
\affiliation{Kavli Institute for Theoretical Physics, University of California, Santa Barbara, California 93106, USA}

\author{Marcos Rigol}
\affiliation{Department of Physics, The Pennsylvania State University, University Park, Pennsylvania 16802, USA}
\affiliation{Kavli Institute for Theoretical Physics, University of California, Santa Barbara, California 93106, USA}

\author{Matthew J. Davis}
\affiliation{The University of Queensland, School of Mathematics and Physics, Brisbane, Queensland 4072, Australia}

\author{Kar\'en V. Kheruntsyan}
\affiliation{The University of Queensland, School of Mathematics and Physics, Brisbane, Queensland 4072, Australia}

\begin{abstract}
We demonstrate the role of interactions in driving the relaxation of an isolated integrable quantum system following a sudden quench.  We consider a family of integrable hard-core lattice anyon models that continuously interpolates between noninteracting spinless fermions and strongly interacting hard-core bosons.  A generalized Jordan-Wigner transformation maps the entire family to noninteracting fermions.  We find that, aside from the singular free-fermion limit, the entire single-particle density matrix and therefore all one-body observables relax to the predictions of the generalized Gibbs ensemble (GGE).  This demonstrates that, in the presence of interactions, correlations between particles in the many-body wave function provide the effective dissipation required to drive relaxation of all one-body observables to the GGE.  This relaxation does not depend on translational invariance, or the tracing out of any spatial domain of the system.
\end{abstract}


\date{\today}

\maketitle
One-dimensional (1D) quantum systems exhibit two features unfamiliar in the three-dimensional world.  The first is the breakdown of the strict distinction between bosonic and fermionic particle statistics~\cite{Girardeau65,Leinaas77}, and the second is the prospect of integrability in the presence of interactions~\cite{Sutherland04}.  Integrable models have been of particular interest as they can be studied using exact analytic and computational approaches to gain insights into strongly correlated quantum systems~\cite{Cazalilla11}.  A recent surge of interest in the nonequilibrium dynamics of these systems \cite{Rigol08,Dziarmaga10,Cazalilla10,Polkovnikov11,Cazalilla11} has been motivated by the failure of some quasi-1D systems in cold-atom experiments~\cite{Kinoshita06,Gring12} to relax to states consistent with conventional statistical mechanics.

A paradigmatic model in this realm is that of lattice hard-core bosons (HCBs), which is integrable by virtue of an exact mapping via the Jordan-Wigner transformation to a system of noninteracting spinless fermions (SFs)~\cite{Jordan28}.  Rigol \emph{et al.}~\cite{Rigol07} showed that, following an abrupt change of Hamiltonian parameters (quantum quench), certain single-particle properties of HCBs such as site and momentum occupations relax to stationary distributions that are not consistent with the predictions of conventional statistical ensembles, but can be described by a generalized Gibbs ensemble (GGE).  The GGE is obtained by maximizing the entropy subject to the constraints that the mean values of the conserved quantities $\hat{I}_\ell$ that make the system integrable are fixed to their values in the initial state.  This yields the density matrix
\begin{equation}\label{eq:rho_GGE}
    \hat{\rho}_{\mathrm{GGE}} = Z_{\mathrm{GGE}}^{-1} \exp\left(-\sum\nolimits_{\ell}\lambda_\ell \hat{I}_\ell\right),
\end{equation}
where the Lagrange multipliers $\lambda_\ell$ are such that $\mathrm{Tr}\{\hat{\rho}_{\mathrm{GGE}}\hat{I}_\ell\} = \langle\hat{I}_\ell\rangle_I$, with $\langle \cdots \rangle_I$ denoting an expectation value taken in the initial (pre-quench) state of the system, and the partition function $Z_\mathrm{GGE} = \mathrm{Tr}\{\exp(-\sum_\ell \lambda_\ell \hat{I}_\ell)\}$.

The validity of the GGE for various classes of observables has now been verified for the relaxed states following quenches of HCBs in a number of distinct geometries~\cite{Rigol06,Rigol07,Cassidy11,Gramsch12} and in a range of other integrable systems~\cite{Cazalilla06,Kollar08,Iucci09,Iucci10,Mossel10,Fioretto10,Calabrese11,Cazalilla12,Caux12,Ziraldo12,Calabrese12,Ziraldo13,Collura13,Caux13,Mussardo13}.  However, the role of interactions in the relaxation dynamics and the true extent of the validity of the GGE as a description of the relaxed state have not been conclusively established.  In particular, recent results have shown that although for SFs the \emph{time-averaged} values of all one-body observables agree with the GGE~\cite{CamposVenuti13,Ziraldo13,He13}, there exist observables that do not \emph{relax} to these mean values, even in the absence of real-space localization~\cite{He13}.  As HCBs can be mapped onto SFs, one is left to wonder if there exist some one-body observables of HCBs that similarly fail to relax.

To elucidate the role of interactions in the relaxation of integrable systems we study the dynamics, following a quench, of a family of hard-core anyons (HCAs) that continuously interpolate between noninteracting SFs and HCBs~\cite{Girardeau06,Averin07,Hao09}.  We find that the \emph{entire} single-particle density matrix relaxes to the GGE prediction for all models in the HCA family, aside from the singular limit of noninteracting SFs.  This implies that the (mixed) state of any particle in the system is driven to a generalized equilibrium by an effective bath formed by the other particles, provided that the particles are interacting.  We contrast this picture with previous works~\cite{Cramer08,Barthel08,Banuls11,Calabrese11,Calabrese12,Fagotti13} that emphasized that relaxation of an isolated quantum system is only observed after tracing out a physical region of the system, which provides dissipation in obvious analogy to the external reservoir traditionally invoked when introducing the \mbox{(grand-)canonical} ensemble.

We note that several proposals for the realization of anyonic statistics in ultracold quantum gas experiments have been made in recent years~\cite{Duan03,Aguado08,Jiang08,Keilmann10}.  Here, we focus on the model of HCAs, which satisfy the generalized commutation relations~\cite{Hao09}
\begin{equation}\label{eq:anyon_comm}
    \hat{a}_j^{}\hat{a}_k^\dagger = \delta_{jk} - e^{-i\theta\,\mathrm{sgn}(j-k)} \hat{a}_k^\dagger \hat{a}_j^{}
    \;\; \text{and} \;\;
    \hat{a}_j^{}\hat{a}_k^{} = - e^{i\theta\,\mathrm{sgn}(j-k)} \hat{a}_k^{}\hat{a}_j^{},
\end{equation}
where the statistical parameter $0\leq \theta \leq \pi$.  When $j=k$, Eq.~\eqref{eq:anyon_comm} yields the hard-core constraints $\hat{a}_j^2=\hat{a}_j^{\dagger 2}=0$ and $\{\hat{a}_j^{},\hat{a}_j^\dagger\}=1$~\cite{Hao09}.  For $\theta=0$ and $\theta=\pi$, Eq.~\eqref{eq:anyon_comm} reduces to the commutation relations of SFs and HCBs, respectively, whereas for $0<\theta<\pi$, these relations interpolate continuously between the two limiting algebras.  For all values of $\theta$, HCAs can be mapped onto SFs by a generalized Jordan-Wigner transformation~\cite{Girardeau06,Hao09,Hao12}.  This makes possible an efficient numerical evaluation of time-evolving one-body observables and their GGE expectation values~\cite{Rigol04b,Rigol05a,Rigol05b,Supp}.

We consider the dynamics of HCAs in a tight-binding model subject to open boundary conditions, with Hamiltonian
\begin{equation}\label{eq:H_TB}
    {\hat H} = -J \sum_{j=1}^{L-1} \left( {\hat a}_j^{\dagger} {\hat a}_{j+1}^{} + \text{H.c.} \right).
\end{equation}
Hereafter, we work in units where the hopping parameter $J=\hbar=1$.  We investigate the dynamics of a system of $N$ particles on a lattice of $L=4N$ sites.  Similarly to Ref.~\cite{Rigol07}, we take as the initial state the ground state of $N$ particles on a smaller sublattice of $2N$ sites, located in the center of the larger lattice (i.e., sites $j=L/4+1,\dots,3L/4$).  At times $t\geq 0$, the HCAs are allowed to freely move in the larger lattice, corresponding to the evolution of the $N$-particle wave function $|\Psi(t)\rangle$ under the action of Hamiltonian~\eqref{eq:H_TB}.  The corresponding GGE is defined by Eq.~\eqref{eq:rho_GGE}, where the $\hat{I}_\ell$ are the occupation numbers of the single-particle energy eigenstates of Hamiltonian~\eqref{eq:H_TB} in the underlying SF model, and $\lambda_\ell = \log[(1-\langle\hat{I}_\ell\rangle_I)/\langle\hat{I}_\ell\rangle_I]$~\cite{Rigol07}.  To characterize the relaxation dynamics of the system, we focus on the properties of the single-particle density matrix $\sigma(t)$~\cite{Blaizot86}, which has elements $\sigma_{jj'}(t) = \langle\Psi(t)| \hat{a}_j^\dagger\hat{a}^{}_{j'}|\Psi(t)\rangle$ in real space.  The corresponding momentum distribution is $m_k(t)=(1/L)\sum_{jj'}e^{ik(j-j')}\sigma_{jj'}(t)$.

\begin{figure}[!t]
    \includegraphics[width=0.485\textwidth]{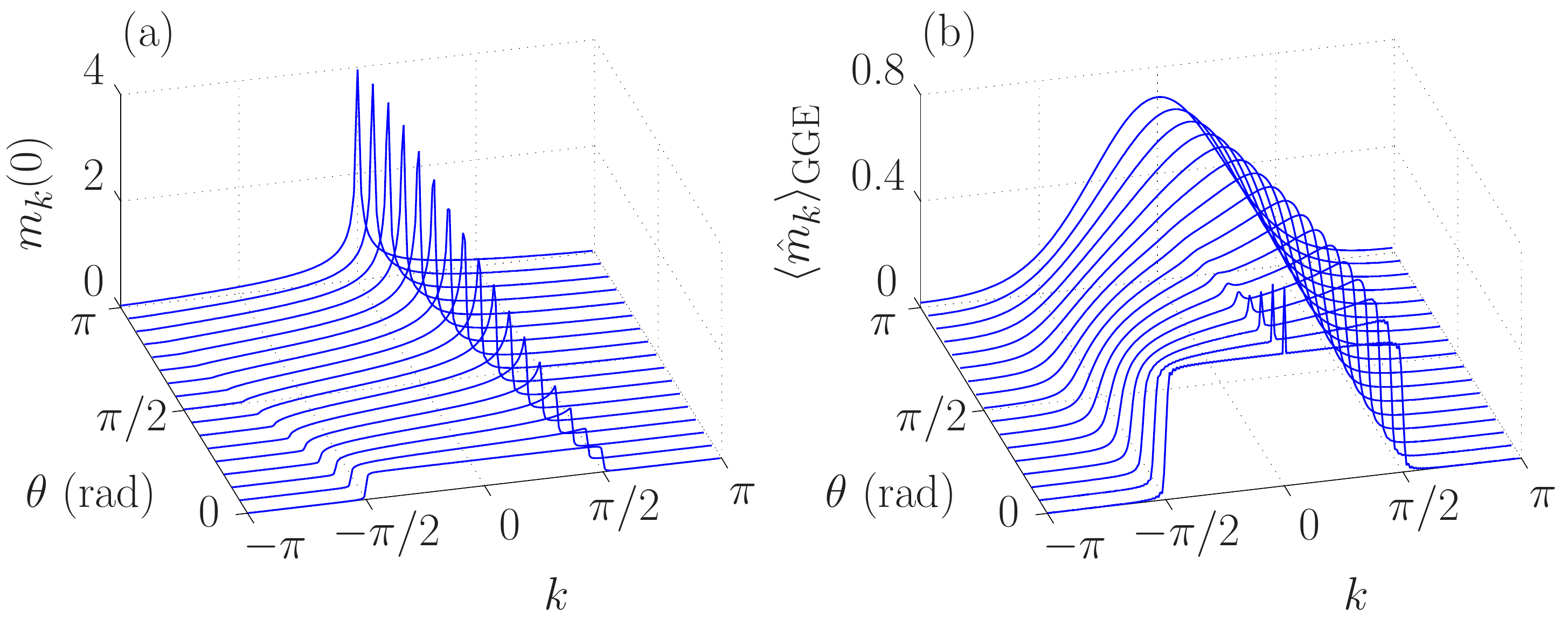}
    \caption{\label{fig:nki_and_nkgge} (Color online) (a) Ground-state momentum distribution of $N=64$ hard-core anyons in a box of $L=128$ sites for various values of the anyon statistical parameter $\theta$.  (b) Momentum distribution of the anyons after (symmetric) expansion into a larger box of $L=256$ lattice sites, as predicted by the GGE (see text).}
\end{figure}

In Fig.~\ref{fig:nki_and_nkgge}(a), we show the initial momentum distribution $m_k(0)$ of HCAs for various values of $\theta$.  For $0<\theta<\pi$, $m_k(0)$ exhibits the characteristic asymmetry of an anyonic state~\cite{Santachiara08,delCampo08,Batchelor08,Hao09}, and interpolates smoothly between the familiar forms of SFs ($\theta=0$) and HCBs ($\theta=\pi$) at zero temperature.  Figure~\ref{fig:nki_and_nkgge}(b) shows the corresponding momentum distributions after relaxation as predicted by the GGE, $\langle\hat{m}_k\rangle_\mathrm{GGE}=(1/L)\sum_{jj'}e^{ik(j-j')}\langle \hat{a}_j^\dagger \hat{a}_{j'}^{} \rangle_\mathrm{GGE}$ (where $\langle \cdots \rangle_\mathrm{GGE} \equiv \mathrm{Tr}\{\hat{\rho}_\mathrm{GGE}\cdots\}$).  We note that even in the SF limit $\langle \hat{m}_k\rangle_\mathrm{GGE}$ is distinct from the initial momentum distribution $m_k(0)$, as the single-particle energy eigenstates of the open-chain Hamiltonian~\eqref{eq:H_TB}~\cite{Busch87} are not momentum eigenmodes.

\begin{figure}
    \includegraphics[width=0.485\textwidth]{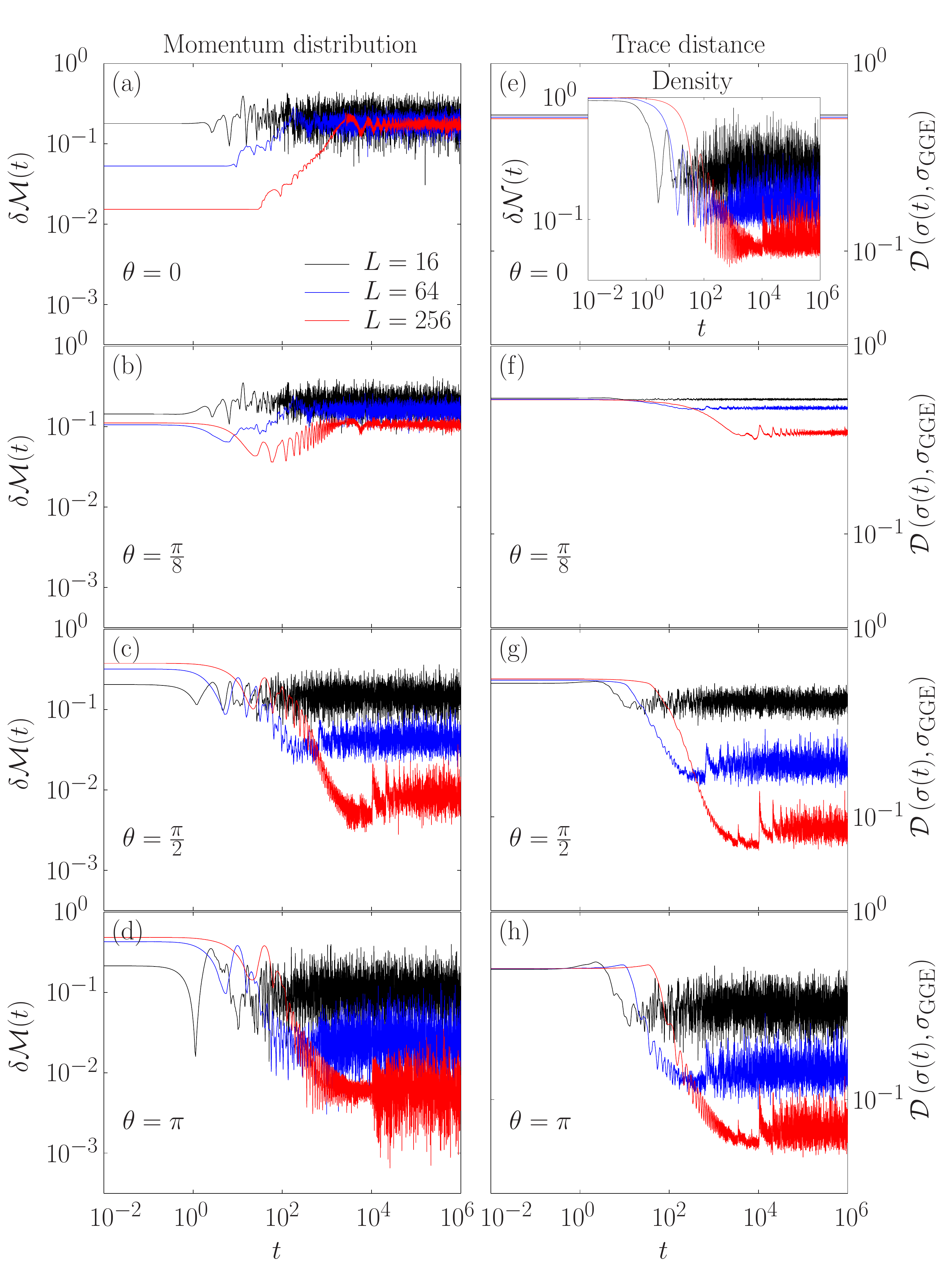}
    \caption{\label{fig:relaxation1} (Color online) Time evolution of (a)--(d) the normalized distance $\delta \mathcal{M}(t)$ between instantaneous and GGE momentum distributions, and (e)--(h) the normalized trace distance $\mathcal{D}(\sigma(t),\sigma_\mathrm{GGE})$, for $N$ hard-core anyons expanding from a hard-wall box of $2N$ sites to one of $L=4N$ sites.  The inset to (e) shows the time evolution of the normalized distance $\delta \mathcal{N}(t)$ between instantaneous and GGE site occupations, which is identical for all statistical parameters $\theta$.}
\end{figure}

The HCA site occupation operators $\hat{n}_j=\hat{a}_j^\dagger \hat{a}_j^{}$ are, for all nonzero values of $\theta$, identical to those of the SF limit ($\theta=0$).  Thus, the time-evolving site occupations $n_j(t) = \sigma_{jj}(t)$ following the quench, and the GGE predictions $\langle \hat{n}_j \rangle_\mathrm{GGE} = \langle \hat{a}_j^\dagger \hat{a}_j^{} \rangle_\mathrm{GGE}$, are common to all cases in the HCA family.  We quantify the difference between $n_j(t)$ and $\langle \hat{n}_j \rangle_\mathrm{GGE}$ by the normalized distance $\delta\mathcal{N}(t)=(\sum_j |n_j(t) - \langle \hat{n}_j\rangle_\mathrm{GGE}|)/ \sum_j\langle \hat{n}_j\rangle_\mathrm{GGE}$ \cite{Gramsch12,He13}, which we plot for three system sizes in the inset to Fig.~\ref{fig:relaxation1}(e).  We see that $\delta \mathcal{N}(t)$ undergoes some initial transient oscillations before decaying to a finite value about which it fluctuates.  This value decreases with increasing system size $L$, suggesting that the site occupations after relaxation converge to the GGE predictions in the limit $L\to\infty$.

In contrast to the site occupations, the time evolution of the momentum distribution $m_k(t)$ depends strongly on $\theta$.  We characterize this evolution by the normalized distance $\delta \mathcal{M}(t) = (\sum_k |m_k(t) - \langle \hat{m}_k \rangle_\mathrm{GGE}|)/\sum_k \langle \hat{m}_k\rangle_\mathrm{GGE}$ between the instantaneous and GGE momentum distributions.  In Figs.~\ref{fig:relaxation1}(a)--\ref{fig:relaxation1}(d), we plot $\delta \mathcal{M}(t)$ for a representative selection of statistical parameters $\theta$.  For HCBs [$\theta=\pi$, shown in Fig.~\ref{fig:relaxation1}(d)] the behavior of $\delta \mathcal{M}(t)$ is similar to that of $\delta \mathcal{N}(t)$, and consistent with previous work~\cite{Rigol07,Gramsch12}.  By contrast, in the SF limit [$\theta=0$, shown in Fig.~\ref{fig:relaxation1}(a)], $\delta \mathcal{M}(t)$ fluctuates about an average value at late times that \emph{does not} decrease significantly with increasing $L$.  In fact, at late times $\delta \mathcal{M}(t)$ is in general \emph{larger} than its value at time $t=0$.  For intermediate values of $\theta$, the behavior of $\delta \mathcal{M}(t)$ interpolates between that seen in the SF and HCB limits, with the late-time values of $\delta \mathcal{M}(t)$ decreasing more significantly with increasing $L$ as $\theta$ departs from the SF limit.

Our results suggest that for any $\theta>0$, just as for HCBs, both $n_j(t)$ and $m_k(t)$ relax to the GGE predictions in the thermodynamic limit, whereas for SFs $n_j(t)$ relaxes to $\langle \hat{n}_j\rangle_\mathrm{GGE}$ but $m_k(t)$ exhibits persistent fluctuations about $\langle\hat{m}_k\rangle_\mathrm{GGE}$~\cite{Gramsch12,He13}.  The absence of relaxation of the SF momentum distribution raises the question of whether there are \emph{some} one-body observables of HCAs with $\theta>0$ that --- in contrast to $n_j$ and $m_k$ --- fail to relax.  To answer this question, we employ a generalization of the distances $\delta \mathcal{M}(t)$ and $\delta \mathcal{N}(t)$ that accounts for \emph{all} one-body observables: the trace distance~\cite{Nielsen00}
\begin{equation}\label{eq:trace_distance}
    \mathcal{D}\left(\sigma(t),\sigma_\mathrm{GGE}\right) =\frac{1}{2N} \mathrm{Tr}\left\{\sqrt{\left(\sigma(t)-\sigma_\mathrm{GGE}\right)^2}\right\}
\end{equation}
between the instantaneous $\sigma(t)$ and the GGE prediction $\sigma_\mathrm{GGE}$.  As both single-particle density matrices have an extensive trace $\mathrm{Tr}\{\sigma(t)\}=\mathrm{Tr}\{\sigma_\mathrm{GGE}\}=N$, we normalize Eq.~\eqref{eq:trace_distance} by $N$, yielding a quantitative measure of any \emph{extensive} difference between $\sigma(t)$ and $\sigma_\mathrm{GGE}$.  The trace distance $\mathcal{D}(\sigma(t),\sigma_\mathrm{GGE})$ thus provides a strict upper bound~\cite{Nielsen00} to the distances $\delta \mathcal{N}(t)$ and $\delta \mathcal{M}(t)$, and indeed the analogous distance $\delta \mathcal{O}(t) = (\sum_q |o_q(t) - \langle \hat{o}_q \rangle_\mathrm{GGE}|)/\sum_q \langle \hat{o}_q \rangle_\mathrm{GGE}$ between the instantaneous occupations $o_q(t)$ of any \emph{arbitrary} complete basis of single-particle modes $|q\rangle$ (e.g., the natural orbitals studied in Refs.~\cite{Rigol06,Gramsch12}) and the GGE predictions $\langle \hat{o}_q \rangle_\mathrm{GGE}$ for these occupations.

In Figs.~\ref{fig:relaxation1}(e)--\ref{fig:relaxation1}(h) we plot the trace distance $\mathcal{D}(\sigma(t),\sigma_\mathrm{GGE})$ and observe that for any $\theta>0$, $\mathcal{D}(\sigma(t),\sigma_\mathrm{GGE})$ decays to an average value about which it fluctuates.  This value decreases with increasing $L$ at fixed $\theta$, and with increasing $\theta$ at fixed $L$.  The decay occurs over a timescale that is similar to that of the initial transient regime seen in $\delta \mathcal{M}(t)$ and $\delta \mathcal{N}(t)$.  However, following the decay, $\mathcal{D}(\sigma(t),\sigma_\mathrm{GGE})$ is \emph{always} lower than the initial value $\mathcal{D}(\sigma(0),\sigma_\mathrm{GGE})$ for all $\theta>0$ and all $L$; i.e., the dynamics of the interacting HCA models always drive the single-particle density matrix closer to the GGE prediction.  For SFs ($\theta=0$), on the other hand, the trace distance is constant in time.  This is a consequence of the invariance of Eq.~\eqref{eq:trace_distance} under unitary transformations and the fact that $\sigma_\mathrm{GGE}$ is diagonal in the single-particle eigenbasis of Hamiltonian~\eqref{eq:H_TB}~\cite{Supp}.

\begin{figure}
    \includegraphics[width=0.485\textwidth]{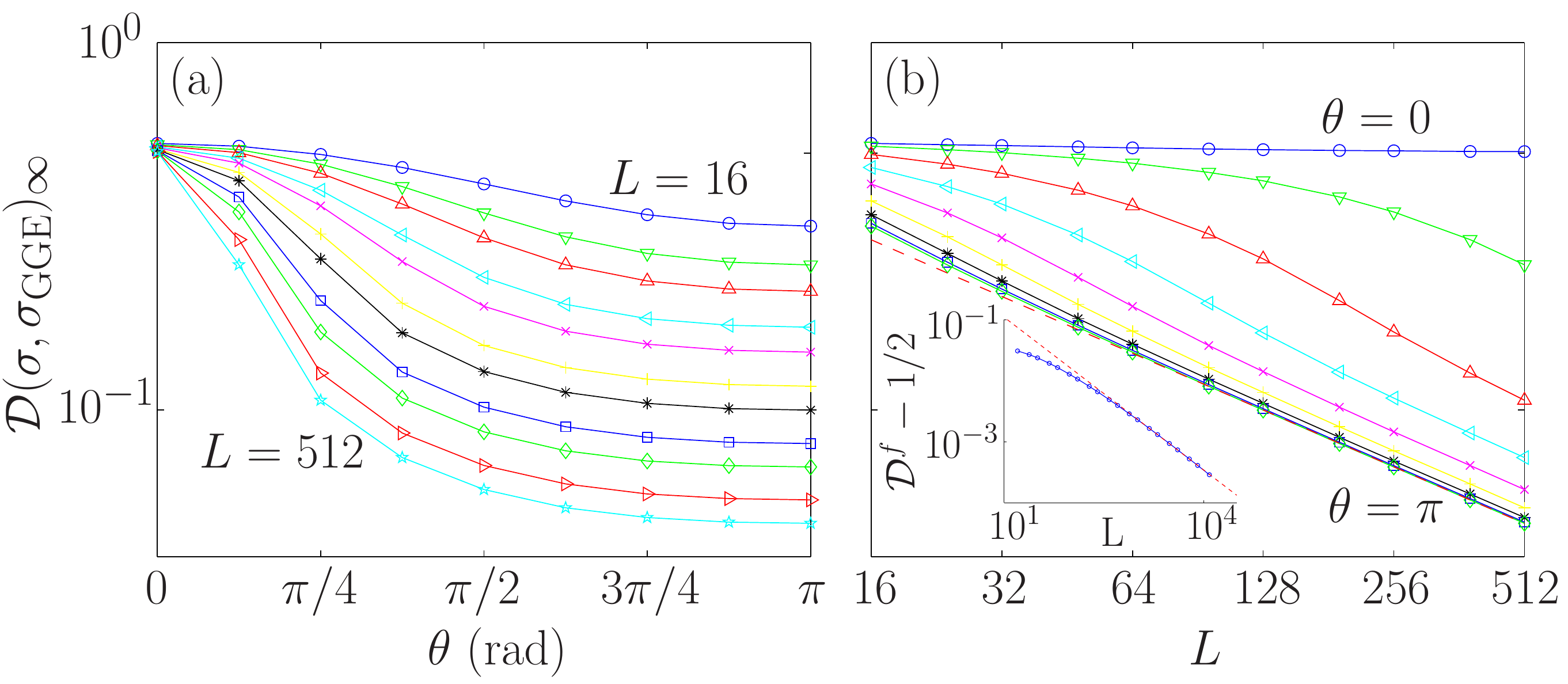}
    \caption{\label{fig:equilibrium_td} (Color online) Dependence of $\mathcal{D}(\sigma,\sigma_\mathrm{GGE})_\infty$ (see text) on (a) the HCA statistical parameter $\theta$ and (b) the system size $L$ (both panels report the same data).  The red dashed line indicates a power-law fit (fitted to data points for $L=96,\dots,512$), which yields $\mathcal{D}(\sigma,\sigma_\mathrm{GGE})_\infty \propto L^{-0.516\pm0.001}$ in the limit of HCBs ($\theta=\pi$).  Inset: Dependence of the SF trace distance $\mathcal{D}^f$ on $L$.  The corresponding dashed line indicates a power-law fit (fitted to data points for $L=1024,\dots,12288$), which yields $\mathcal{D}^f-1/2 \propto L^{-0.833\pm0.004}$.}
\end{figure}

In order to quantitatively characterize the convergence of $\sigma(t)$ to $\sigma_\mathrm{GGE}$, we consider the time average of $\mathcal{D}(\sigma(t),\sigma_\mathrm{GGE})$ over the period $t\in [10^5,10^6]$ (i.e., after relaxation) and denote this quantity by $\mathcal{D}(\sigma,\sigma_\mathrm{GGE})_\infty$.  Figure~\ref{fig:equilibrium_td}(a) shows the dependence of $\mathcal{D}(\sigma,\sigma_\mathrm{GGE})_\infty$ on the statistical parameter $\theta$ for a selection of system sizes $L=16,24,32,48,\dots,512$.  The results make it clear that the time-averaged trace distance decreases with increasing $\theta$, and that its initial decrease as $\theta$ is increased from zero becomes steeper with increasing $L$.  The latter suggests that any $\theta>0$ leads to a vanishing trace distance between $\sigma(t)$ after relaxation and $\sigma_\mathrm{GGE}$ in the thermodynamic limit.  In Fig.~\ref{fig:equilibrium_td}(b) we present the same data as a function of system size.  We observe that the decrease in $\mathcal{D}(\sigma,\sigma_\mathrm{GGE})_\infty$ with increasing $L$ is more pronounced for larger values of $\theta$, whereas for SFs there is little change in $\mathcal{D}(\sigma,\sigma_\mathrm{GGE})_\infty$ with increasing system size.

For HCBs ($\theta=\pi$), a fit to the data reveals that the time-averaged trace distance after relaxation exhibits power-law scaling close to $L^{-1/2}$.  For all interacting models ($\theta>0$) $\mathcal{D}(\sigma,\sigma_\mathrm{GGE})_\infty$ appears to exhibit an identical scaling at large $L$, although systems with smaller values of $\theta$ are slower to reach this limiting behavior.  By contrast, for the case of SFs the (time-invariant) trace distance, which we denote by $\mathcal{D}^f$, does not appear to scale towards zero as $L\to\infty$.  In fact, a nonvanishing lower bound $\mathcal{D}^f \geq 1/2$ can be derived in the thermodynamic limit~\cite{Supp}.  An examination of $\mathcal{D}^f$ for larger lattice sizes $L$ [inset to Fig.~\ref{fig:equilibrium_td}(b)] suggests that this bound is in fact an equality in the thermodynamic limit, as $\mathcal{D}^f-1/2$ appears to vanish as a power law (close to $L^{-5/6}$) as $L\to\infty$.

The behavior of $\mathcal{D}(\sigma,\sigma_\mathrm{GGE})_\infty$ for HCA models with $\theta>0$ implies that the entire single-particle density matrix, and as a result the occupations of \emph{all} single-particle bases (not just the site and momentum occupations), relax to the predictions of the GGE in the thermodynamic limit.  This occurs in spite of the persistence of time fluctuations of one-body observables in the underlying SF model~\cite{Ziraldo13,He13}.  Indeed, the vanishing of $\mathcal{D}(\sigma,\sigma_\mathrm{GGE})_\infty$ with increasing system size implies that no measurement of one-body observables can reveal any more than a sub-extensive distinction between the state after relaxation and the GGE prediction~\cite{Supp}.  This is in stark contrast to the behavior of the underlying SFs, for which the time-evolving $\sigma(t)$ can always (in principle) be distinguished from $\sigma_\mathrm{GGE}$.

In the quenches we have considered so far, the final Hamiltonian $\hat{H}$ lacks translational invariance only because of the imposed boundary conditions, the effects of which one might expect to vanish in the thermodynamic limit.  It is therefore natural to ask whether the complete relaxation of the anyonic single-particle density matrices we have observed applies only to translationally invariant systems.  To answer this question we follow Refs.~\cite{Gramsch12,He13} in explicitly breaking translational invariance by adding to Hamiltonian~\eqref{eq:H_TB} a lattice potential $\hat{V}_\mathrm{lat} = \lambda \sum_{j=1}^L \cos (2 \pi \varsigma j) \hat{a}_j^\dagger\hat{a}_j^{}$ with an incommensurate period $1/\varsigma = 2/(\sqrt{5}-1)$, which yields the Aubry-Andr\'e (AA) Hamiltonian~\cite{Aubry80}.  We then repeat the expansion quench, considering now the evolution of $N$ HCAs under the action of the AA Hamiltonian with $\lambda=1$ (for which the single-particle energy eigenstates remain delocalized) starting from the ground state on the central $L/2$ sites of the AA superlattice.

\begin{figure}
    \includegraphics[width=0.30\textwidth]{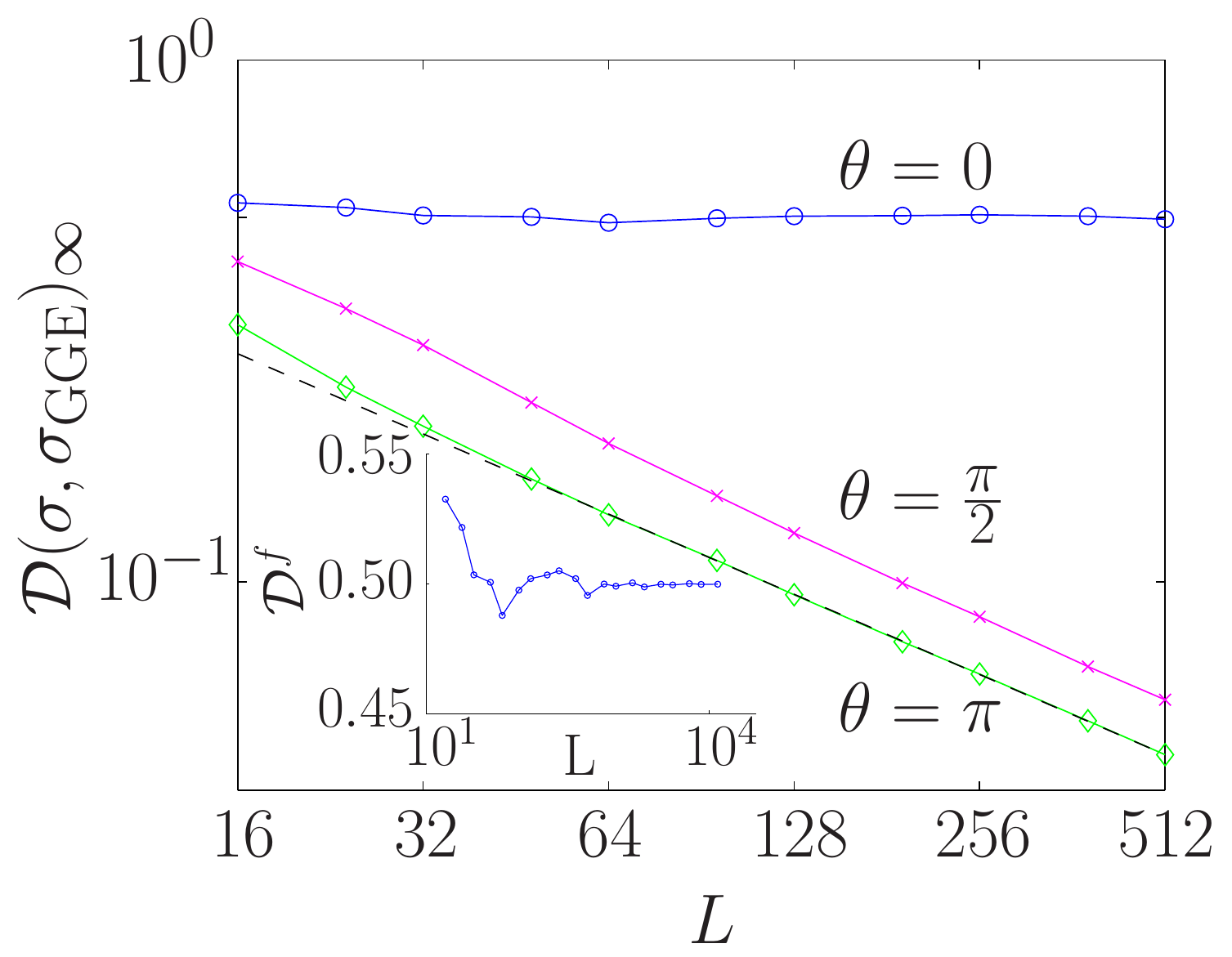}
    \caption{\label{fig:equilibrium_td_aa} (Color online) Dependence of $\mathcal{D}(\sigma,\sigma_\mathrm{GGE})_\infty$ on $L$, for HCAs undergoing expansion within the Aubry-Andr\'e model with $\lambda = 1$.  The black dashed line indicates a power-law fit (fitted to data points for $L=96,\dots,512$), which yields $\mathcal{D}(\sigma,\sigma_\mathrm{GGE})_\infty \propto L^{-0.512\pm0.003}$ for HCBs ($\theta=\pi$).  Inset: Dependence of the SF trace distance $\mathcal{D}^f$ on $L$.}
\end{figure}

In Fig.~\ref{fig:equilibrium_td_aa}, we plot the time-averaged trace distance $\mathcal{D}(\sigma,\sigma_\mathrm{GGE})_\infty$ following expansion in the AA model as a function of system size for $\theta=0,\ \pi/2,$ and $\pi$.  In the limit of SFs, $\mathcal{D}^f$ does not decay significantly with increasing $L$, and, as shown in the inset, appears to saturate at a value $\mathcal{D}^f\approx 1/2$.  By contrast, the trace distances $\mathcal{D}(\sigma,\sigma_\mathrm{GGE})_\infty$ of the interacting models of hard-core semions ($\theta=\pi/2$) and bosons ($\theta = \pi$) exhibit a clear power-law scaling close to $L^{-1/2}$.  These results demonstrate that translational invariance of the final Hamiltonian (in the limit $L\to\infty$) is not required for the post-quench single-particle density matrices of HCAs (and HCBs in particular) to relax to $\sigma_\mathrm{GGE}$ in the thermodynamic limit.  We note, however, that real-space localization (as produced, e.g., by the AA Hamiltonian for $\lambda>2$) would necessarily preclude relaxation of $\sigma(t)$ to the GGE, as the resulting saturation of the time average of $\delta \mathcal{N}(t)$~\cite{Gramsch12,He13} bounds $\mathcal{D}(\sigma,\sigma_\mathrm{GGE})_\infty$ to be nonzero for all values of the statistical parameter $\theta$~\cite{Supp}.

We emphasize that this complete one-body relaxation of interacting HCAs is in stark contrast to the behavior of noninteracting SFs.  The results presented here, and in Refs.~\cite{Gramsch12,Ziraldo13,He13}, demonstrate the existence of extensive sets of one-body SF observables that exhibit non-vanishing time fluctuations in the thermodynamic limit, whereas the fact that $\mathcal{D}(\sigma,\sigma_\mathrm{GGE})_\infty$ vanishes with increasing $L$ for $\theta>0$ implies that no such extensive set exists in the interacting HCA models~\cite{Supp}.  We note also that, for the interacting quenches considered here~\cite{Supp} and in Ref.~\cite{Gramsch12}, $\delta \mathcal{M}(t)$ scales like $L^{-1}$ as $L\to\infty$.  Although this scaling is in fact \emph{faster} than the $L^{-1/2}$ scaling observed for relaxing SF observables in Ref.~\cite{He13}, the trace distance $\mathcal{D}\left(\sigma,\sigma_\mathrm{GGE}\right)_\infty$ scales like $L^{-1/2}$.  Remarkably, the characteristic scaling of the Gaussian equilibration scenario proposed for free SF models in Ref.~\cite{CamposVenuti13}, although violated for certain observables of the SF model itself, is in effect restored by the generalized Jordan-Wigner transformation to an interacting hard-core anyon/boson model when the entire single-particle density matrix is considered.

In summary, we have shown that a family of HCAs that interpolates between SFs and HCBs exhibits complete relaxation of the entire single-particle density matrix to the GGE prediction, with the singular exception of the noninteracting SF model itself.  This demonstrates that relaxation to the GGE can manifest without tracing out any spatial domain of the system; i.e., that the dissipation required for the single-particle state to relax to (a generalized) equilibrium is provided by the correlations between particles in the many-body wave function~\cite{Santachiara07}.  Interactions are, however, seen to play a crucial role in this relaxation:  Although for noninteracting SFs the single-particle density matrix agrees with the GGE after time averaging~\cite{Ziraldo13,He13}, it is precluded from relaxing by the unitarity of its evolution.  By contrast, for all interacting HCA models (with $\theta>0$) the single-particle density matrix evolves non-unitarily and each particle in the system is driven, by an effective bath provided by the other particles, to the stationary state predicted by the GGE.  We have also shown that these conclusions apply even when translational invariance of the pre- and post-quench Hamiltonians is broken, provided that the system remains delocalized.

\begin{acknowledgments}
This work was supported by the Australian Research Council Discovery Project~DP110101047 (T.W., M.D., K.K.), the Office of Naval Research (M.R.), and the National Science Foundation under Grant No.~PHY11-25915 (T.W., M.R.).
\end{acknowledgments}

\bibliographystyle{prsty}

\end{document}


\title{Supplementary Material --- Nonequilibrium dynamics of one-dimensional hard-core anyons following a quench: Complete relaxation of one-body observables}

\author{Tod M. Wright}
\affiliation{The University of Queensland, School of Mathematics and Physics, Brisbane, Queensland 4072, Australia}
\affiliation{Kavli Institute for Theoretical Physics, University of California, Santa Barbara, California 93106, USA}

\author{Marcos Rigol}
\affiliation{Department of Physics, The Pennsylvania State University, University Park, Pennsylvania 16802, USA}
\affiliation{Kavli Institute for Theoretical Physics, University of California, Santa Barbara, California 93106, USA}

\author{Matthew J. Davis}
\affiliation{The University of Queensland, School of Mathematics and Physics, Brisbane, Queensland 4072, Australia}

\author{Kar\'en V. Kheruntsyan}
\affiliation{The University of Queensland, School of Mathematics and Physics, Brisbane, Queensland 4072, Australia}

\date{\today}
\maketitle

\renewcommand{\thepage}{S\arabic{page}}

\section{Calculation of anyonic observables}
\vspace{-0.5\baselineskip}
Hard-core anyons, with commutation relations
\begin{align}\label{eq:anyon_comm}
    \hat{a}_j^{}\hat{a}_k^\dagger &= \delta_{jk} - e^{-i\theta\,\mathrm{sgn}(j-k)} \hat{a}_k^\dagger \hat{a}_j^{}, \nonumber \\
    \hat{a}_j^{}\hat{a}_k^{} &= - e^{i\theta\,\mathrm{sgn}(j-k)} \hat{a}_k^{}\hat{a}_j^{},
\end{align}
can be mapped to spinless fermions ($\{\hat{f}_j^{},\hat{f}_k^{}\}=0$, $\{\hat{f}_j^{},\hat{f}_k^\dagger\}=\delta_{jk}$) by the generalized Jordan-Wigner transformation
\begin{align}\label{eq:jordan_wigner}
    \hat{a}_j^\dagger &= \hat{f}_j^\dagger \prod_{k<j} e^{i\theta \hat{f}_k^\dagger \hat{f}_k^{}}, \quad
    \hat{a}_j^{} = \prod_{k<j} e^{-i\theta \hat{f}_k^\dagger \hat{f}_k^{}} \hat{f}_j^{},
\end{align}
and in particular, in the limit $\theta=0$, $\hat{a}_j^{(\dagger)}=\hat{f}_j^{(\dagger)}$.  One-body observables in a pure state $|\Psi_A\rangle$ of HCAs can thus be calculated by a simple generalization of the method of Ref.~\cite{Rigol04b}.  The HCA Green's function can be expressed as an SF expectation value
\begin{align}
    G_{ij} &\equiv \langle \Psi_A | \hat{a}_i^{} \hat{a}_j^\dagger | \Psi_A \rangle \nonumber \\
    &= \langle \Psi_F |\prod_{k<i} e^{-i\theta \hat{f}_k^\dagger \hat{f}_k^{}} \hat{f}_i^{}  \hat{f}_j^\dagger \prod_{l<j} e^{i\theta \hat{f}_l^\dagger \hat{f}_l^{}} |\Psi_F \rangle
\end{align}
in the Slater determinant $|\Psi_F \rangle = \prod_{m=1}^N \sum_{n=1}^L P_{nm} \hat{f}_n^\dagger |0\rangle$.  Therefore
\begin{align}
    \hat{f}_j^\dagger \prod_{k<j} e^{i\theta \hat{f}_k^\dagger \hat{f}_k^{}} |\Psi_F\rangle &= \prod_{m=1}^{N+1} \sum_{n=1}^L P_{nm}^j \hat{f}_n^\dagger |0\rangle,
\end{align}
where $P_{nm}^j$ is obtained from $P_{nm}$ by multiplying the elements $m<j$ by the phase $e^{i\theta}$, and the addition of one column, with the sole nonzero element $P_{jN+1}=1$.  Thus
\begin{align}
   G_{ij} &= \langle 0 | \prod_{m=1}^{N+1} \sum_{n=1}^L (P_{nm}^i)^*\, \hat{f}_n \prod_{k=1}^{N+1}\sum_{l=1}^L P_{lk}^j \hat{f}_l^\dagger |0\rangle \nonumber \\
    &= \mathrm{det}\left[ (\mathbf{P}^i)^\dagger \mathbf{P}^j \right].
\end{align}
The single-particle density matrix of HCAs is then given by
\begin{align}
    \sigma_{ij} &\equiv \langle \hat{a}_i^\dagger \hat{a}_j^{} \rangle = \delta_{ij} -e^{-i\theta \mathrm{sgn}(i-j)} G_{ij}.
\end{align}
This methodology can be straightforwardly extended to compute the time evolution of the nonequilibrium single-particle density matrix after a quench, following the approach described in Ref.~\cite{Rigol05a}.

The single-particle density matrix in the GGE $(\sigma_\mathrm{GGE})_{ij} = Z_\mathrm{GGE}^{-1} \mathrm{Tr}\{\hat{a}_i^\dagger \hat{a}_j^{} e^{-\sum_{\ell=1}^L \lambda_\ell \hat{I}_\ell} \}$, where $\hat{I}_\ell \equiv \tilde{f}_\ell^\dagger\tilde{f}_\ell^{}$ are the occupations of SFs in the appropriate single-particle orbitals $|\zeta_\ell\rangle$, can similarly be evaluated by an extension of the method of Ref.~\cite{Rigol05b}.  In particular, using Eq.~\eqref{eq:jordan_wigner} we have
\begin{align}
    (\sigma_\mathrm{GGE})_{ij} &= \frac{1}{Z_\mathrm{GGE}} \mathrm{Tr}\Bigg\{\hat{f}_i^\dagger \hat{f}_j^{} \prod_{k=1}^{j-1} e^{-i\theta \hat{f}_k^\dagger\hat{f}_k^{}} \\ 
    &\qquad\qquad\qquad \times e^{-\sum_{\ell=1}^L \lambda_\ell \hat{I}_\ell} \prod_{l=1}^{i-1} e^{i\theta \hat{f}_l^\dagger\hat{f}_l^{}}\Bigg\},\nonumber 
\end{align}
which, for $i\neq j$, can be expressed (cf. Ref.~\cite{Rigol05b})
\begin{align}
    (\sigma_\mathrm{GGE})_{ij} &= Z_\mathrm{GGE}^{-1} \Big\{ \mathrm{det}\left[\mathbf{I} + (\mathbf{I}+\mathbf{A}^{ij})\mathbf{O}_1\mathbf{U}
    e^{-\bm{\lambda}}\mathbf{U}^\dagger\mathbf{O}_2\right] \nonumber \\
    &\qquad\qquad\;\; - \mathrm{det}\left[\mathbf{I} + \mathbf{O}_1 \mathbf{U} e^{-\bm{\lambda}}\mathbf{U}^\dagger\mathbf{O}_2\right]\Big\},
\end{align}
where $\mathbf{I}$ is the $L\times L$ identity matrix, $\mathbf{A}^{ij}$ is such that $(\mathbf{A}^{ij})_{i'j'}=\delta_{ii'}\delta_{jj'}$, $\mathbf{O}_1$ ($\mathbf{O}_2$) is diagonal with the first $i$ ($j$) diagonal elements equal to $e^{-i\theta}$ ($e^{i\theta}$) and the others equal to unity, $\mathbf{U}$ is the unitary transformation to the diagonal representation of the $|\zeta_\ell\rangle$, and $\bm{\lambda}$ is diagonal with nonzero elements $(\bm{\lambda})_{\ell\ell} = \lambda_\ell$.  The GGE partition function $Z_\mathrm{GGE} = \prod_\ell (1 + e^{-\lambda_\ell})$, and the diagonal elements of the GGE single-particle density matrix $(\sigma_\mathrm{GGE})_{ii} = Z_\mathrm{GGE}^{-1}[\mathbf{U} (\mathbf{I} + e^{-\bm{\lambda}} )^{-1} \mathbf{U}^\dagger]_{ii}$.

\section{Single-particle trace distance}
We define the trace distance between the single-particle density matrices $\sigma(t)$ and $\sigma_\mathrm{GGE}$
\begin{align}\label{eq:trace_distance_supp}
    \mathcal{D}(\sigma(t),\sigma_\mathrm{GGE}) &=\frac{1}{2N} \mathrm{Tr}\left\{\sqrt{\left(\sigma(t)-\sigma_\mathrm{GGE}\right)^2}\right\}.
\end{align}
This differs from the standard definition~\cite{Nielsen00} of the trace distance between normalized density matrices by a factor $\mathrm{Tr}\{\sigma(t)\}=\mathrm{Tr}\{\sigma_\mathrm{GGE}\}=N=\nu L$, with $\nu$ the filling fraction.  Standard results for the trace distance immediately transpose to the definition of Eq.~\eqref{eq:trace_distance_supp}.  In particular, the normalized distance \mbox{$\delta \mathcal{O}(t) = (\sum_q |o_q(t) - \langle \hat{o}_q \rangle_\mathrm{GGE}|)/\sum_q \langle \hat{o}_q \rangle_\mathrm{GGE}$} between the instantaneous occupations $o_q(t)=\langle q|\sigma(t)|q\rangle$ of any arbitrary complete basis of single-particle modes $|q\rangle$ (e.g., site or momentum modes, or the natural orbitals of $\sigma_\mathrm{GGE}$) and the GGE predictions $\langle \hat{o}_q \rangle_\mathrm{GGE}=\langle q|\sigma_\mathrm{GGE}|q\rangle$ for these occupations, satisfies $\delta \mathcal{O}(t) \leq 2 \mathcal{D}(\sigma(t),\sigma_\mathrm{GGE})$.  It follows therefore that if the time-averaged trace distance $\mathcal{D}(\sigma,\sigma_\mathrm{GGE})_\infty$ exhibits power-law decay like $L^{-\alpha}$, then the time average $\delta \mathcal{O}_\infty$ of $\delta \mathcal{O}(t)$ after relaxation scales to zero at least as rapidly.  Conversely, the existence of any such set of modes $|q\rangle$ for which $\delta \mathcal{O}_\infty \sim L^0$ implies that $\mathcal{D}(\sigma(t),\sigma_\mathrm{GGE}) \sim L^0$ also.

Moreover, $\mathcal{D}(\sigma(t),\sigma_\mathrm{GGE})$ is equal to the standard trace distance between the unit-normalized density matrices $\sigma'(t)=\sigma(t)/N$ and $\sigma'_\mathrm{GGE}=\sigma_\mathrm{GGE}/N$.  Thus we have $0 \leq \mathcal{D}(\sigma(t),\sigma_\mathrm{GGE})\leq1$, where the lower bound is reached when $\sigma'(t)=\sigma'_\mathrm{GGE}$; i.e., $\mathcal{D}(\sigma(t),\sigma_\mathrm{GGE})$ vanishes as $L\to\infty$ if and only if $\sigma(t)$ becomes equal to $\sigma_\mathrm{GGE}$, up to subextensive differences in this limit.  More specifically, given any positive operator-valued measure $\{E_m\}$ composed of positive operators $E_m$ acting in the single-particle space such that $\sum_m E_m = {1}$~\cite{Nielsen00}, the distance
\begin{align}
    d(p_m(t),p_m^\mathrm{GGE}) &= \frac{1}{2}\sum_m |\mathrm{Tr}\{E_m(\sigma(t)-\sigma_\mathrm{GGE})\}|
\end{align}
between the distributions of the outcomes $p_m(t)=\mathrm{Tr}\{E_m\sigma(t)\}$ and $p_m^\mathrm{GGE}=\mathrm{Tr}\{E_m\sigma_\mathrm{GGE}\}$ of the measurement defined by $\{E_m\}$ is bounded
\begin{align}
    d(p_m(t),p_m^\mathrm{GGE}) &\leq N \mathcal{D}(\sigma(t),\sigma_\mathrm{GGE}) \sim \nu L^{1-\alpha},
\end{align}
if $\mathcal{D}(\sigma(t),\sigma_\mathrm{GGE})\sim L^{-\alpha}$.  Thus if $\alpha>0$ at most only a subextensive difference between $\sigma(t)$ and $\sigma_\mathrm{GGE}$ can be revealed by any measurement in the single-particle space.

\subsection*{Bounds on single-particle trace distance for free spinless fermions}\label{subsec:TD_sat}
We consider an initial state $|\Psi_I\rangle$ that is an eigenstate of a free spinless fermion Hamiltonian $\hat{H}_I= a^\dagger h_I a$, where $a^\dagger = (\hat{a}_1^\dagger,\dots,\hat{a}_L^\dagger)$; i.e., $|\Psi_I\rangle$ is a Slater determinant of $N$ single-particle energy eigenstates $|\chi_\ell\rangle$ of $h_I$.  At times $t>0$, $|\Psi(t)\rangle$ evolves under the action of an SF Hamiltonian $\hat{H}_F = a^\dagger h_F a$, with single-particle energy eigenvectors $|\zeta_j\rangle$ ($h_F|\zeta_j\rangle=\epsilon_j|\zeta_j\rangle$).  The resulting time-dependent Slater determinant $|\Psi(t)\rangle$ is completely characterized (up to a global phase) by the single-particle density matrix~\cite{Blaizot86}
\begin{align}
    \sigma(t) &= U(t)\sigma(0)U^\dagger(t) \nonumber \\
    &= \sum_{j,k=1}^L \langle \zeta_j | \sigma(0) |\zeta_k \rangle  e^{-i(\epsilon_j - \epsilon_k)t/\hbar} |\zeta_j\rangle\langle\zeta_{k}|,
\end{align}
where the time-evolution operator $U(t) = \exp(-i h_F t/\hbar)$, and $\sigma(0) = \sum_\ell |\chi_\ell\rangle\langle\chi_\ell|$.  We compare this single-particle density matrix to the single-particle density matrix of the generalized Gibbs ensemble,
\begin{equation}
    \sigma_\mathrm{GGE} = \mathrm{Tr}_{2\cdots N}\{\hat{\rho}_\mathrm{GGE}\} = \sum_{j=1}^L \overline{n}_j |\zeta_j\rangle \langle \zeta_j|,
\end{equation}
with
\begin{equation}
    \overline{n}_j = \langle \Psi_I | \tilde{a}_j^\dagger \tilde{a}_j^{} |\Psi_I\rangle = \mathrm{Tr}\{\sigma(0) |\zeta_j\rangle\langle\zeta_j|\},
\end{equation}
where $\tilde{a}_j^{}$ annihilates an SF in the orbital $|\zeta_j\rangle$.  We note in particular that $\sum_{j=1}^L \overline{n}_j = \mathrm{Tr}\{\sigma(0) \sum_{j=1}^L |\zeta_j\rangle\langle\zeta_j|\} = \mathrm{Tr}\{\sigma(0)\}=N$, and that $U(t)\sigma_\mathrm{GGE}U^\dagger(t)=\sigma_\mathrm{GGE}$, as $\sigma_\mathrm{GGE}$ is diagonal in the basis $\{|\zeta_j\rangle\}$.

We consider now the normalized trace distance~\eqref{eq:trace_distance_supp} between $\sigma(t)$ and $\sigma_\mathrm{GGE}$.  Standard results for the trace distance~\cite{Nielsen00} imply that
\begin{align}
    \mathcal{D}(\sigma(t),\sigma_\mathrm{GGE}) &= \mathcal{D}(U(t)\sigma(0) U^\dagger(t),U(t) \sigma_\mathrm{GGE} U^\dagger(t)) \nonumber \\
    &= \mathcal{D}(\sigma(0),\sigma_\mathrm{GGE});
\end{align}
i.e., the trace distance is invariant under the evolution of $\sigma(t)$, and that
\begin{equation}\label{eq:proj_bound}
    \mathcal{D}(\sigma(0),\sigma_\mathrm{GGE}) \geq \frac{1}{N}\mathrm{Tr}\{\mathcal{P} (\sigma_\mathrm{GGE} - \sigma(0)) \},
\end{equation}
where $\mathcal{P}$ is any projector in the single-particle space spanned by $\{|\zeta_j\rangle\}$.  Considering in particular the projector $\mathcal{Q} = {1} - \sigma(0)$ we have therefore
\begin{align}
    \mathcal{D}(\sigma(0),\sigma_\mathrm{GGE})
    &\geq\frac{1}{N}\mathrm{Tr}\{\sigma_\mathrm{GGE} - \sigma(0)\sigma_\mathrm{GGE}\}.
\end{align}
As $\sigma_\mathrm{GGE}$ is diagonal in the basis $\{|\zeta_j\rangle\}$,
\begin{align}
    \mathrm{Tr}\{\sigma(0)\,\sigma_\mathrm{GGE}\}
    &= \sum_{j=1}^L \langle\zeta_j|\sigma(0) |\zeta_{j}\rangle\langle\zeta_{j}| \sigma_\mathrm{GGE} |\zeta_j\rangle \nonumber \\
    &= \mathrm{Tr}\{(\sigma_\mathrm{GGE})^2\},
\end{align}
and so we obtain a lower bound
\begin{align}\label{eq:purity_bound}
    \mathcal{D}(\sigma(0),\sigma_\mathrm{GGE}) &\geq 1 - \gamma_\mathrm{GGE}
\end{align}
for the trace distance in terms of the generalized purity $\gamma_\mathrm{GGE} \equiv (1/N)\mathrm{Tr}\{(\sigma_\mathrm{GGE})^2\}$.
\\

A necessary condition for $\mathcal{D}(\sigma(0),\sigma_\mathrm{GGE})$ to vanish in the thermodynamic limit is that $\gamma_\mathrm{GGE} \to 1$ as $L\to\infty$; i.e., that $\sigma_\mathrm{GGE}$ becomes idempotent as $L\to\infty$.  Conversely, an idempotent $\sigma_\mathrm{GGE}$ is the single-particle density matrix of some single Slater determinant~\cite{Blaizot86}
\begin{equation}
    \sigma_\mathrm{GGE} = \mathrm{Tr}_{2\cdots N}\{|\Xi\rangle\langle\Xi|\},
\end{equation}
where $|\Xi\rangle$ is composed of $N$ orbitals $|\zeta_{j_\ell}\rangle$, for $\ell=1,\dots,N$.  In this limit $\langle\zeta_{j_\ell} |\sigma(0)|\zeta_{j_\ell}\rangle = 1$; i.e., $\sigma(0)$ is an identity operator within the space spanned by the $N$ orbitals $|\zeta_{j_\ell}\rangle$, and thus the spans of $\{|\chi_j\rangle\}$ and $\{|\zeta_{j_\ell}\rangle\}$ are equal.  This occurs in particular if the initial and final Hamiltonians are related by a unitary transformation $h_F = U h_I U^\dagger$ (or if $\nu=1$), in which case the individual single-particle orbitals $|\chi_\ell(t)\rangle = e^{-ih_Ft}|\chi_\ell\rangle$ evolve nontrivially in time, but all observables are time-independent.  In fact, the invariance of a Slater determinant under unitary transformations among the orbitals of which it is composed~\cite{Blaizot86} implies that $|\Xi\rangle$ is equal to $|\Psi_I\rangle$ (up to a phase); i.e., $|\Psi_I\rangle$ is an energy eigenstate of $\hat{H}_F$, and the diagonal-ensemble density matrix~\cite{Rigol08} is pure: $\hat{\rho}_\mathrm{DE} \equiv |\Psi_I\rangle\langle\Psi_I|$.  In this limit $\sigma_\mathrm{GGE} = \mathrm{Tr}_{2\cdots N}\{\hat{\rho}_\mathrm{DE}\}=\sigma(0)$, and therefore in general $\mathcal{D}(\sigma(0),\sigma_\mathrm{GGE})\to 0$ if and only if the diagonal-ensemble purity $\mathrm{\Gamma}_\mathrm{DE}\equiv\mathrm{Tr}\{(\hat{\rho}_\mathrm{DE})^2\} \to 1$ in the thermodynamic limit (a trivial ``quench'').
\\

\begin{figure}
    \includegraphics[width=0.485\textwidth]{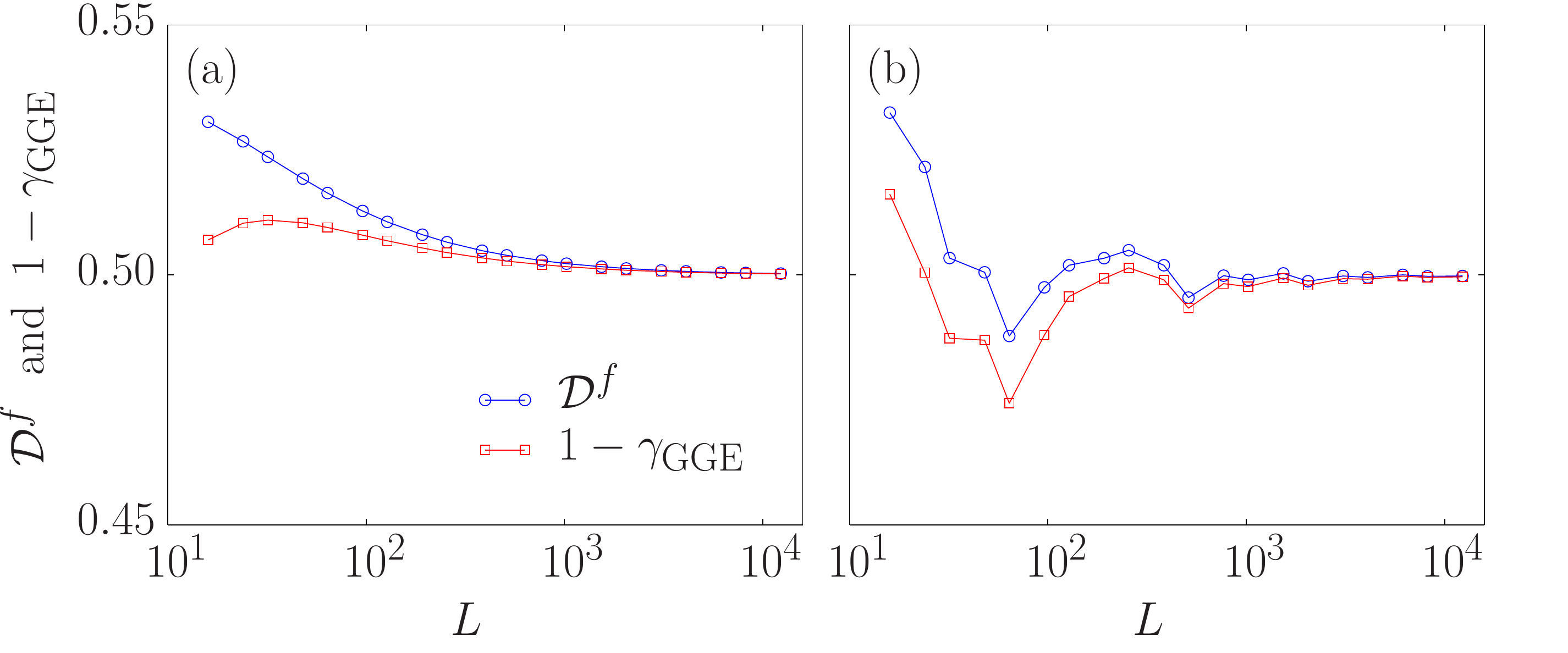}
    \caption{\label{fig:tdf_vs_tr} (Color online) Comparison of $\mathcal{D}^f$, the (time-independent) trace distance between the instantaneous single-particle density matrix $\sigma(t)$ of SFs and the corresponding GGE prediction, to the bound provided by the generalized purity $\gamma_\mathrm{GGE}$ of $\sigma_\mathrm{GGE}$ (see text).  Results are shown for expansion quenches in (a) the tight-binding model (b) the Aubry-Andr\'e model.}
\end{figure}

In Fig.~\ref{fig:tdf_vs_tr}, we compare the bound of Eq.~\eqref{eq:purity_bound} with the numerically calculated trace distance $\mathcal{D}^f$ between $\sigma(0)$ and $\sigma_\mathrm{GGE}$ for SFs, for the expansion quenches considered in the main text.  We note that in general $\mathcal{D}^f$ is somewhat larger than the bound of Eq.~\eqref{eq:purity_bound} but, both without [Fig.~\ref{fig:tdf_vs_tr}(a)] and with [Fig.~\ref{fig:tdf_vs_tr}(b)] the additional incommensurate lattice potential, the results suggest that (for these quenches) this bound becomes an equality and, indeed, that $\mathcal{D}^f\to 1/2$ in the thermodynamic limit.
\\

In fact, this asymptotic value of $\mathcal{D}^f$ has a simple physical interpretation for the expansion quenches.  In the quench within the tight-binding model, the GGE site occupations $\langle \hat{n}_j \rangle_\mathrm{GGE}$ become uniform as $L\to\infty$.  Thus, introducing a projector $\mathcal{P} = \sum_{j\in L\backslash I} |j\rangle\langle j|$, where $L\backslash I$ denotes the set of lattice sites outside the central box in which the particles are confined before the quench, we have $\mathrm{Tr}\{\mathcal{P}(\sigma_\mathrm{GGE}-\sigma(0))\}/N = \mathrm{Tr}\{\mathcal{P}\sigma_\mathrm{GGE}\}/N \to 1/2$ as $L\to\infty$, bounding $\mathcal{D}^f \geq 1/2$ in this limit.  In the quench within the delocalized regime of the AA model, the GGE site occupations exhibit persistent site-to-site fluctuations (cf. Ref.~\cite{Gramsch12}).  Nevertheless, one expects the average site occupancy in any extensive block of contiguous sites to converge to $1/2$ in the thermodynamic limit, implying the same asymptotic behavior $\mathrm{Tr}\{\mathcal{P}\sigma_\mathrm{GGE}\}/N \to 1/2$ as $L\to\infty$.  Thus in both quenches the trace distance $1/2$ between $\sigma(t)$ and $\sigma_\mathrm{GGE}$ in the thermodynamic limit can be understood as the ``memory'' of the initial confinement of the sample within the smaller box, which is additional to the initial-state memory encoded in the conserved quantities $\hat{I}_\ell$ that define the GGE.  This memory is preserved by the unitarity of the evolution in the single-particle space in the limit of SFs, but lost in the case of the interacting HCA models, for which $\sigma(t)$ evolves non-unitarily; i.e., in this case the single-particle density matrix is effectively coupled, due to correlations in the $N$-body wave function, to a bath that induces relaxation to a generalized equilibrium state constrained only by the mean values of the conserved quantities $\hat{I}_\ell$.

\section{Time fluctuations of observables}
\vspace{-1.5\baselineskip}
\subsection*{Free spinless fermions}
\vspace{-1.0\baselineskip}
The fact that $\mathcal{D}(\sigma(t),\sigma_\mathrm{GGE})$ has a time-invariant, nonvanishing value in the SF limit implies that at each given time $t$ the single-particle density matrix $\sigma(t)$ can be distinguished from $\sigma_\mathrm{GGE}$.  It does not necessarily imply the existence of a time-independent single-particle basis $|q\rangle$ for which the occupations $o_q(t)$ have nonvanishing fluctuations (i.e., $\delta \mathcal{O}_\infty \sim L^0$).

A general one-body (i.e., quadratic) observable has the form $\hat{O}=\sum_{ij=1}^L \hat{a}_i^\dagger o_{ij} \hat{a}_j^{} = a^\dagger o a$, where $o$ is an $L\times L$ Hermitian matrix.  Such observables therefore form a vector space, corresponding to the dimension-$L^2$ space of complex Hermitian matrices.  An orthonormal [with respect to the Hilbert-Schmidt inner product $(A|B) \equiv \mathrm{Tr}\{A^\dagger B\}$] basis for this space is given, in terms of the elementary $L\times L$ matrices $[E^{kl}]_{ij}=\delta_{ik}\delta_{jl}$, by the set of Hermitian matrices $E^{jj}$ for $1\leq j \leq L$, together with $(E^{jk}+E^{kj})/\sqrt{2}$ and $i(E^{jk}-E^{kj})/\sqrt{2}$ for $1 \leq j < k \leq L$.  We therefore consider a basis set of observables $\hat{O}_\ell = a^\dagger o_\ell a$, where $o_\ell$ is an arbitrary enumeration of the aforementioned $L^2$ basis matrices.  Introducing the notation $\overline{f} \equiv \lim_{T\to\infty} (1/T) \int_0^T\! dt\, f(t)$, the time average and variance of each observable are given by $\overline{O}_\ell = \langle \hat{O}_\ell \rangle_\mathrm{GGE}$ and
\begin{align}
    \Delta_\ell^2 &\equiv \overline{ \left( O_\ell(t) - \overline{O}_\ell \right)^2 } =
    \overline{\mathrm{Tr}\{o_\ell (\sigma(t)-\sigma_\mathrm{GGE})\}^2},
\end{align}
respectively.  Averaging the variance over all $L^2$ observables yields
\begin{align}
    \Delta^2 &\equiv \frac{1}{L^2} \sum_{\ell=1}^{L^2} \Delta_\ell^2
    = \frac{1}{L^2} \overline{\sum_{\ell=1}^{L^2} \mathrm{Tr}\{o_\ell (\sigma(t)-\sigma_\mathrm{GGE})\}^2}.
\end{align}
From the definitions of the basis matrices $o_\ell$, we find
\begin{align}\label{eq:hs_identity}
    \sum_{\ell=1}^{L^2} \mathrm{Tr}\{o_\ell (\sigma(t)-\sigma_\mathrm{GGE})\}^2&=\sum_{j,k=1}^{L}|\langle j |\sigma(t)-\sigma_\mathrm{GGE}|k\rangle|^2 \nonumber \\ 
    &= (\sigma(t)-\sigma_\mathrm{GGE}|\sigma(t)-\sigma_\mathrm{GGE}) \nonumber \\
    &= \mathrm{Tr}\{\sigma(0) -\sigma_\mathrm{GGE}^2\},
\end{align}
where we have used the invariance of the Hilbert-Schmidt inner product under unitary transformations.  Therefore,
\begin{align}\label{eq:var_basis}
    \Delta^2 &= \frac{\nu}{L}\left(1 - \gamma_\mathrm{GGE}\right);  
\end{align}
i.e., for non-trivial quenches $\gamma_\mathrm{GGE} < 1$, the variances of one-body SF observables scale to zero as $L^{-1}$, when averaged over a \emph{complete} basis of such observables.  We stress that Eq.~\eqref{eq:var_basis} does not, however, preclude the existence of an extensive number ($\sim L$) of one-body SF observable variances that saturate ($\Delta_\ell \sim L^0$)~\cite{Gramsch12,Ziraldo13,He13}.

\subsection*{Post-quench momentum-distribution fluctuations}
\begin{figure}
    \includegraphics[width=0.485\textwidth]{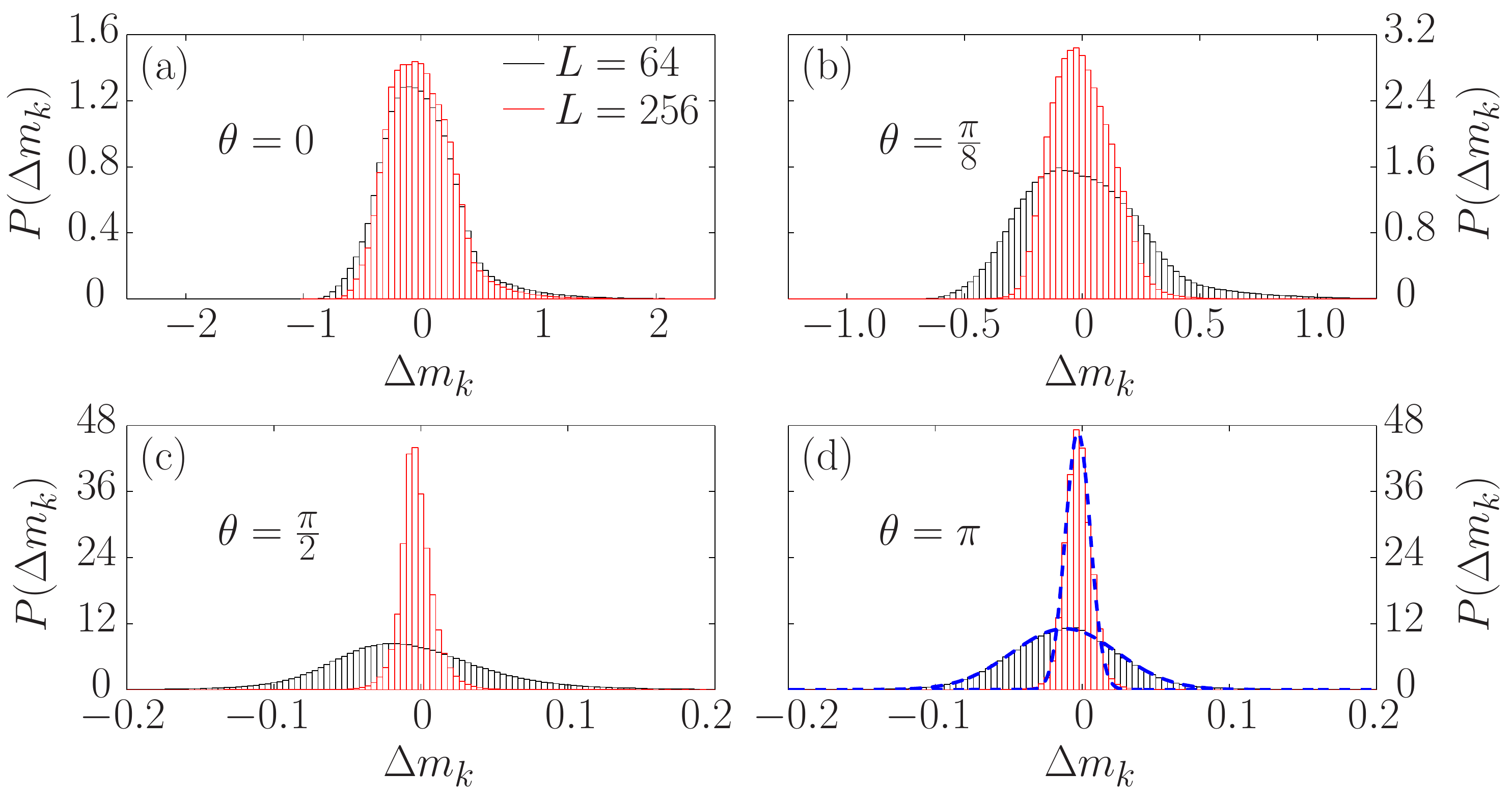}
    \caption{\label{fig:histos} (Color online) Histograms of the normalized time fluctuations $\Delta m_k(t) = [m_k(t) - \langle \hat{m}_k\rangle_\mathrm{GGE}]/\langle \hat{m}_k \rangle_\mathrm{GGE}$ of momentum-mode occupancies following the free expansion quench.  Results are shown for (a) $\theta=0$ (SFs), (b) $\theta=\pi/8$, (c) $\theta=\pi/2$ (hard-core semions), and (d) $\theta=\pi$ (HCBs).  Each histogram corresponds to the distribution of $\Delta m_k(t)$ for all $k$ modes, over times sampled from the period $t\in [10^5,10^6]$.  Dashed lines in (d) indicate Gaussian fits, which yield means $-0.012$ and $-0.003$, and standard deviations $0.0360$ and $0.0085$, for $L=64$ and $L=256$, respectively.}
\end{figure}

To characterize the fluctuations of the HCA momentum distributions following the expansion quench in the tight-binding model, we consider the normalized difference
\begin{align}\label{eq:Delta_mk}
    \Delta m_k(t) = \frac{m_k(t) - \langle \hat{m}_k\rangle_\mathrm{GGE}}{\langle \hat{m}_k \rangle_\mathrm{GGE}}
\end{align}
between the occupation of an \emph{individual} momentum mode $k$ at time $t$ and its mean value in the GGE.  In Fig.~\ref{fig:histos} we plot histograms of the distributions $P(\Delta m_k)$ composed from the values of $\Delta m_k(t)$ for all momenta $k$, sampled from a range of times $t\in [10^5,10^6]$.

A qualitative distinction between the behavior of the time fluctuations of $m_k(t)$ in the limit of SFs [Fig.~\ref{fig:histos}(a)] and that observed for the interacting models with $\theta>0$ [Figs.~\ref{fig:histos}(b)--\ref{fig:histos}(d)] is immediately apparent.  The distribution of fluctuations for SFs is strongly non-Gaussian, with a large tail appearing for positive values of $\Delta m_k(t)$.  It also does not exhibit significant narrowing upon increasing the system size $L$ from $64$ to $256$.  This behavior is consistent with the non-Gaussian, non-narrowing distribution of momentum fluctuations observed in Ref.~\cite{He13} following a quench to a delocalized fermion Hamiltonian\footnote{In Eq.~\eqref{eq:Delta_mk} we normalize each individual difference $\Delta m_k$ by the GGE occupation $\langle \hat{m}_k \rangle_\mathrm{GGE}$ to obtain a representation of the relative fluctuations in each $k$ mode, as in general many such modes are weakly occupied [see Fig.~1(b) in the main text].  We note, therefore, that the time-fluctuation distributions in Fig.~\ref{fig:histos} cannot be directly compared to those of Ref.~\cite{He13}.}.  For the system sizes studied, the histograms of $P(\Delta m_k)$ for quenches of the HCA models become less non-Gaussian as $\theta$ is increased, and in the case of HCBs [Fig.~\ref{fig:histos}(d)] the distributions are very close to Gaussian, as indicated by the fits (dashed lines) to these distributions.  In clear contrast to the case of SFs, for each nonzero $\theta$ the distribution $P(\Delta m_k)$ exhibits significant narrowing as the system size is increased from $L=64$ to $L=256$, suggesting that the time fluctuations of $m_k(t)$ in these quenches become vanishingly small as $L\to\infty$.

\begin{figure}
    \includegraphics[width=0.485\textwidth]{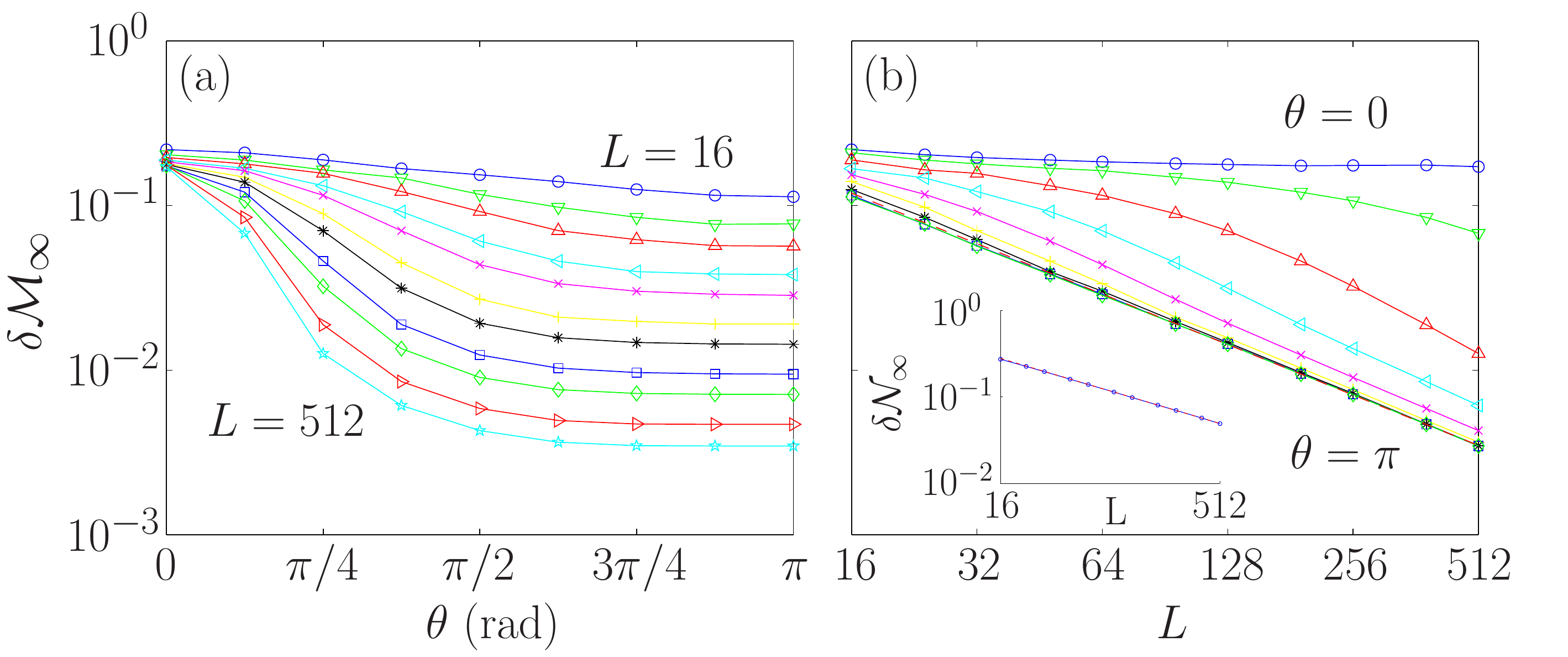}
    \caption{\label{fig:equilibrium_dmk} (Color online) Dependence of $\delta \mathcal{M}_\infty$ (see text) on (a) the HCA statistical parameter $\theta$ and (b) the system size $L$.  The red dashed line indicates a power-law fit (fitted to data points for $L=96,\dots,512$), which yields $\delta \mathcal{M}_\infty \propto L^{-1.018\pm0.005}$ in the limit of HCBs ($\theta=\pi$).  Inset: Dependence of $\delta \mathcal{N}_\infty$ on $L$.  The corresponding dashed line indicates a power-law fit (fitted to data points for $L=96,\dots,512$), which yields $\delta \mathcal{N}_\infty \propto L^{-0.500\pm0.002}$.}
\end{figure}

From the fits to the histograms for $\theta=\pi$, we find that upon increasing the system size from $L=64$ to $L=256$, both the (absolute value of the) mean and standard deviation of the histogram decrease by a factor of approximately four, consistent with $\sim L^{-1}$ convergence of the momentum distribution to the GGE prediction (cf. Ref.~\cite{Gramsch12}).  This scaling is confirmed in Fig.~\ref{fig:equilibrium_dmk}, where we plot the time average of $\delta \mathcal{M}(t)$ over the period $t \in [10^5,10^6]$, which we denote by $\delta \mathcal{M}_\infty$.  In the limit of SFs ($\theta=0$), $\delta \mathcal{M}_\infty$ does not decay appreciably with increasing $L$, whereas in the opposite limit of HCBs ($\theta=\pi$), we observe a scaling close to $\delta \mathcal{M}_\infty \sim L^{-1}$.  The inset to Fig.~\ref{fig:equilibrium_dmk}(b) shows the corresponding time-averaged distance $\delta \mathcal{N}_\infty$ between $n_j(t)$ and $\langle \hat{n}_j \rangle_\mathrm{GGE}$, which is common to all values of $\theta$, and scales like $L^{-1/2}$.

\begin{figure}
    \includegraphics[width=0.485\textwidth]{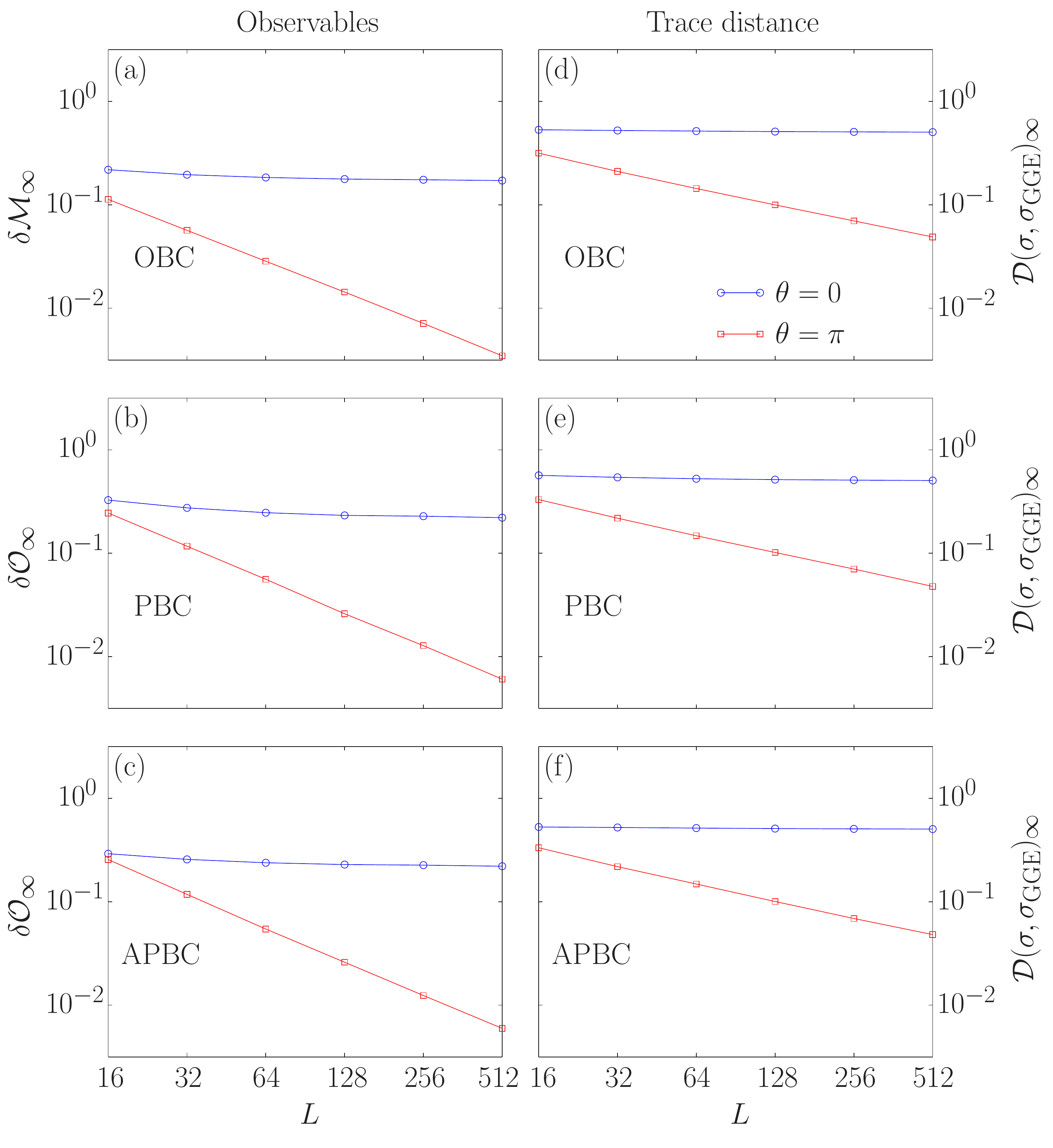}
    \caption{\label{fig:BCs} (Color online) Role of boundary conditions in relaxation in the SF and HCB limits.  The left column shows the time-averaged normalized distance (a) $\delta \mathcal{M}_\infty$ between the post-quench momentum distribution $m_k(t)$ and the GGE prediction $\langle \hat{m}_k\rangle_\mathrm{GGE}$ for a quench to OBCs, (b) $\delta \mathcal{O}_\infty$ between occupations $o_\ell(t)$ of open-boundary single-particle eigenstates and the GGE prediction $\langle \hat{o}_\ell\rangle_\mathrm{GGE}$ for a quench to PBCs, and (c) $\delta \mathcal{O}_\infty$ for a quench to APBCs.  The right column shows the time-averaged normalized trace distance $\mathcal{D}(\sigma,\sigma_\mathrm{GGE})_\infty$ following quenches to (d) OBCs, (e) PBCs, and (f) APBCs.  All time averages correspond to the period $t\in [10^5,10^6]$, and all distances for HCBs are relative to the GGE calculated for a quench to OBCs (see text).}
\end{figure}

\subsection*{Localization and relaxation}
The presence of nonrelaxing variables in the SF model demonstrates a fundamental difference in relaxation behavior between the SF limit, and the remaining models of the HCA family.  Equations~\eqref{eq:hs_identity}~and~\eqref{eq:var_basis} show that for SFs the single-particle density matrix $\sigma(t)$ converges to $\sigma_\mathrm{GGE}$ as $L\to\infty$ in the sense that the normalized Hilbert-Schmidt distance
\begin{align}
    \mathcal{D}_\mathrm{HS}(\sigma(t),\sigma_\mathrm{GGE}) &\equiv \frac{1}{N} \sqrt{\mathrm{Tr}\{(\sigma(t)-\sigma_\mathrm{GGE})^2\}}
\end{align}
between the two vanishes like $L^{-1/2}$ (cf. the Gaussian equilibration scenario of Ref.~\cite{CamposVenuti13}).  This weak convergence, enforced by the unitarity of the evolution of $\sigma(t)$, still allows for the presence of extensive sets of nonrelaxing one-body observables.  By contrast, the stronger trace-distance convergence of the HCA single-particle density matrices to the GGE (also $\sim L^{-1/2}$) implies that no such nonrelaxing sets are present in the interacting models.

From the arguments of Ziraldo~and~Santoro~\cite{Ziraldo13}, we might expect that one can always contrive a set of nonrelaxing observables in an SF quench, by constructing a basis that is localized in (but distinct from) the single-particle energy eigenstates of the post-quench Hamiltonian.  This expectation is strongly supported by the observation of the exotic $L^{-1/4}$ convergence of both $n_j$ and $m_k$ to the GGE in quenches of SFs to the critical (localization/delocalization) point of the AA model~\cite{He13}.

As a further partial check on the role of single-particle localization in precluding relaxation of SF observables and the efficacy of the Jordan-Wigner transformation to an interacting model in removing these persistent fluctuations, we investigate the effects of the boundary conditions of the post-quench Hamiltonian in expansion quenches within the tight-binding model.  In addition to the open boundary conditions (OBCs) considered in the main text, we perform quenches in which the final Hamiltonian has periodic boundary conditions (PBCs) $\hat{a}_{L+1}=\hat{a}_1$ and antiperiodic boundary conditions \mbox{(APBCs)} $\hat{a}_{L+1}=-\hat{a}_1$.  In each case we take as the initial state the ground state of $L/4$ particles on a lattice with $L/2$ sites and open boundary conditions located in the center of the final chain.  For (A)PBCs, the single-particle eigenstates of the final Hamiltonian are plane waves, and thus the SF momentum distribution $m_k(t)=\langle \hat{m}_k\rangle_\mathrm{GGE}$ at all times.  For quenches to (A)PBCs, we therefore consider the occupations $o_\ell(t) = \langle \phi_\ell| \sigma(t) | \phi_\ell \rangle$ of the single-particle energy eigenstates $|\phi_\ell\rangle$ of the tight-binding Hamiltonian with \emph{open} boundary conditions (see, e.g., Ref.~\cite{Busch87}), which are localized in momentum space.  We characterize the relaxation dynamics of these observables by the normalized distance $\delta \mathcal{O}(t) = (\sum_\ell |o_\ell(t) - \langle \hat{o}_\ell \rangle_\mathrm{GGE}|)/ \sum_\ell \langle \hat{o}_\ell \rangle_\mathrm{GGE}$, the time average (over $t\in [10^5,10^6]$) of which we denote by $\delta \mathcal{O}_\infty$ and plot for SFs ($\theta=0$) in PBCs and APBCs in Figs.~\ref{fig:BCs}(b)~and~\ref{fig:BCs}(c) respectively.  Our results show that this distance saturates in both cases, similarly to $\delta \mathcal{M}_\infty$ for SFs in the quench with OBCs [Fig.~\ref{fig:BCs}(a)].  In all three cases, the trace distance $\mathcal{D}^f$ for SFs [Figs.~\ref{fig:BCs}(d)--\ref{fig:BCs}(f)] saturates to $\approx 1/2$.

Turning our attention to the opposite limit of HCBs ($\theta=\pi$), we note that for even particle numbers $N$ such as we consider here, the Jordan-Wigner transformation maps HCBs with PBCs (APBCs) to SFs with APBCs (PBCs), whereas for odd particle numbers HCBs are always mapped to SFs with the same boundary conditions.  As the particle number $N$ is a fluctuating quantity in the GGE~\cite{Rigol05b}, this alternation of boundary conditions precludes us from calculating GGE expectation values for HCBs in (A)PBCs.  For HCBs, we therefore calculate $\delta \mathcal{O}(t)$ and $\mathcal{D}(\sigma,\sigma_\mathrm{GGE})_\infty$ relative to $\sigma_\mathrm{GGE}$ calculated for the system with \emph{open} boundary conditions.  Figures~\ref{fig:BCs}(d)--\ref{fig:BCs}(f) indicate that, although the particular SF observables in which persistent fluctuations manifest depend in general on the boundary conditions, the spurious choice of boundary conditions used in calculating $\sigma_\mathrm{GGE}$ for HCBs becomes irrelevant as $L\rightarrow\infty$, as the trace distance between $\sigma(t)$ and this form of $\sigma_\mathrm{GGE}$ vanishes like $L^{-1/2}$ in all three cases.  Moreover, Figs.~\ref{fig:BCs}(b)~and~\ref{fig:BCs}(c) show that in (A)PBCs $\delta \mathcal{O}_\infty$, although saturating for SFs, vanishes like $L^{-1}$ for HCBs, echoing the behavior of $\delta \mathcal{M}_\infty$ in the original (open) quench geometry [Fig.~\ref{fig:BCs}(a)].

\bibliographystyle{prsty}